\documentclass[conference]{IEEEtran}
\usepackage{cite}
\usepackage{filecontents}
\usepackage{stfloats}
\usepackage{color}

\ifCLASSINFOpdf
  \usepackage[pdftex]{graphicx}
  \graphicspath{{images/}}
  \DeclareGraphicsExtensions{.pdf,.jpeg,.png}
\else
  \usepackage[dvips]{graphicx}
  \graphicspath{{images/}}
  \DeclareGraphicsExtensions{.eps}
\fi

\usepackage[cmex10]{amsmath}

\usepackage{array}

\usepackage{mdwmath}
\usepackage{mdwtab}

\usepackage[tight,footnotesize]{subfigure}
\usepackage{url}
\newtheorem{theorem}{Theorem}
\newtheorem{corollary}{Corollary}
\newtheorem{Lemma}{Lemma}

\usepackage{amssymb}
\usepackage{bm}
\usepackage{mathrsfs}
\newtheorem{remark}{Remark}
\newtheorem{definition}{Definition}

 \def\gap{0.64ex}
   \abovedisplayskip\gap
   \belowdisplayskip\gap
   \abovedisplayshortskip\gap
   \belowdisplayshortskip\gap

\begin{document}
\IEEEoverridecommandlockouts
\title{A New Index Coding Scheme Exploiting Interlinked Cycles}

\author{\IEEEauthorblockN{Chandra Thapa, Lawrence Ong, and Sarah J. Johnson}
\IEEEauthorblockA{School of Electrical Engineering and Computer Science, The University of Newcastle, Newcastle, Australia\\
Email: chandra.thapa@uon.edu.au, lawrence.ong@cantab.net, sarah.johnson@newcastle.edu.au} %
\thanks{This work is supported by the Australian Research Council under
grants FT110100195, FT140100219, and DP150100903.} 
}
\maketitle
\begin{abstract}
We study the index coding problem in the unicast message setting, i.e., where each message is requested by one unique receiver. This problem can be modeled by a directed graph.  We propose a new scheme called interlinked cycle cover, which exploits interlinked cycles in the directed graph, for designing index codes. This new scheme generalizes the existing clique cover and cycle cover schemes. We prove that for a class of infinitely many digraphs with messages of any length, interlinked cycle cover provides an optimal index code. Furthermore, the index code is linear with linear time encoding complexity.
\end{abstract}

\begin{IEEEkeywords}
Index coding, unicast, optimal broadcast rate, linear codes, interlinked cycles.
\end{IEEEkeywords}

%
\IEEEpeerreviewmaketitle

\section{Introduction}
We consider a source sending message packets through a noiseless broadcast channel to multiple receivers, each knowing some packets a priori, which is known as \emph{side information}. One can exploit the \emph{side information} to reduce the number of coded packets to be sent by the source, for all receivers to decode their requested messages. This is known as the index coding problem and was introduced by Birk and Kol in 1998~\cite{ISCOD}. The problem can be modeled by a digraph (i.e., directed graph). The aim is to find an optimal scheme, which provides the minimum number of coded packets. Birk and Kol used graph theory to find upper and lower bounds to the minimum number of coded packets. Subsequently, tighter bounds were found using various approaches including graph theory~\cite{ISCOD,maisbound,neely,chaudhary,localgarphcoloring}, Shannon random coding~\cite{composite,unalwagner13}, numerical approaches, i.e., linear programming~\cite{linearprogramming}, and interference alignment~\cite{interferencealignment,topologicalinterferencemanagement}. However, the index coding problem remains open to date. 

Among graph-theoretic approaches, clique cover~\cite{ISCOD} and cycle cover~\cite{maisbound,neely,chaudhary} are useful as they provide insights on how to code on specific graph structures (as opposed to numerical approaches) and they are valid for message packets of any length (as opposed to random coding). However, they code on disjoint cycles and cliques on the digraph, ignoring useful side information captured in interlinked cycles.  In this paper, we propose a new scheme, called interlinked cycle cover ($ \mathsf{ICC} $), to exploit interlinked cycles. The $\mathsf{ICC}$ scheme turns out to be a generalization of clique cover and cycle cover. 

Index codes generated by $\mathsf{ICC}$ are scalar linear codes. Linear codes simplify encoding and decoding process over non-linear codes. Ong~\cite{linearcodeoptimal} \cite{newlinearcodeoptimal}, found some classes of graphs where scalar linear codes are optimal. These classes of graphs have either five vertices or fewer, or the property that the removal of two vertices results in a maximum acyclic induced subgraph (MAIS). In fact, optimal linear codes for a digraph can be found using the minrank function \cite{maisbound}, which is, however, NP-hard to compute~\cite{NPhard1} in general. In this paper, we characterize a class of digraphs for which 
scalar linear codes generated by $\mathsf{ICC}$ are optimal.


\subsection{Our Contributions}
\begin{enumerate}
\item We propose a new index coding scheme, $ \mathsf{ICC} $, which  generalizes the cycle cover and the clique cover schemes.
\item We show that for some digraphs, $ \mathsf{ICC} $  can outperform existing techniques for message packets of  finite length.
\item We characterize a class of digraphs where $ \mathsf{ICC} $ is optimal (over all codes, including non-linear index codes).
\end{enumerate}

\section{Definitions}
Suppose we have an index coding problem in which a source wants to send $ n $ message packets $ X = \{x_1,x_2,\dotsc,x_n\} $ to $ n $ receivers, where each receiver is requesting a unique message packet $ x_i $ (i.e., unicast), and each receiver has some side information, $ S_i\subseteq X\setminus \{x_i\} $. This problem can be described by a digraph $D =(V,A) $, where $ V=\{v_1,v_2,\dotsc,v_n\} $ is the set of vertices representing the $ n $ receivers. An arc $ (v_i\rightarrow v_j) \in A $ exists from vertex $ v_i $ to vertex $ v_j$ if receiver $ v_i $ has packet $ x_j $ (requested by receiver $ v_j $) as its side information. If vertex $ v_i $ has an out-neighborhood $ N_D^+ (v_i) $, then the side information of $ v_i $ is $ S_i = \{x_j: v_j\in N_D^+(v_i)\} $. For simplicity, we use the term ``messages'' to refer to message packets in the remainder of this paper. 

\begin{definition}
\emph{(Valid index codes)} Let $ x_i \in \{0,1\}^t $ for all $ i $, and for some integer $ t\geq 1 $, i.e., each message contains $t$ binary bits. 
Given an index coding problem $D$, a valid index code ($ \mathscr{F} $,$ \{\mathscr{G}_i\} $) is defined as follows:
\begin{enumerate}
\item An encoding function for the source, $ \mathscr{F}:\{0,1\}^{nt} \rightarrow \{0,1\}^p $, which maps $X$ to a $ p $-bit index for some integer $ p $. 
\item A decoding function $ \mathscr{G}_i $ for every receiver $ v_i $, $ \mathscr{G}_i: \{0,1\}^p \times \{0,1\}^{|S_i|t}\rightarrow \{0,1\}^t $, that maps the received index $ \mathscr{F}(X) $ and its side information $ S_i $ to the requested  message $ x_i $.   
\end{enumerate}  

 The broadcast rate of the ($ \mathscr{F},\{\mathscr{G}_i\} $) index code is the number of transmitted bits per received message bits at every user, or the number of coded packets (of $t$ bits), denoted by $ \ell_t(D) \triangleq \frac{p}{t}$. Thus, the optimal broadcast rate for a given index coding problem $D$ with  $t$-bit messages is $ \beta_t(D)=\underset{\mathscr{F}}{\mathrm{min}}\ \frac{p}{t}=\underset{\mathscr{F}}{\mathrm{min}}\ \ell_t(D) \triangleq \ell_t^*(D) $. 
For a given index coding problem $D$, the minimum optimal broadcast rate over all $ t $ is defined as $ \beta(D)=\underset{t}{\mathrm{inf}}\ \beta_t(D)$.  
\end{definition}
\begin{definition}
\emph{(Path and cycle)} In a digraph, a \emph{path} comprises a sequence of distinct (except possibly the first and last) vertices, say $u_1,u_2,\dotsc,u_L $, and, an arc $( u_i \rightarrow u_{i+1} )$ for each consecutive pair of vertices $ (u_i, u_{i+1}) $ for all $ i\in \{1,\dotsc,L-1\} $.  Here, $u_1$ is called the \emph{initial} vertex, and $u_L$ the \emph{terminal} vertex of the path.  If the initial vertex and terminal vertex of a path are the same, then it is called a \emph{cycle}. 
\end{definition}  




\begin{figure}[t!]
 		\includegraphics[scale=0.5]{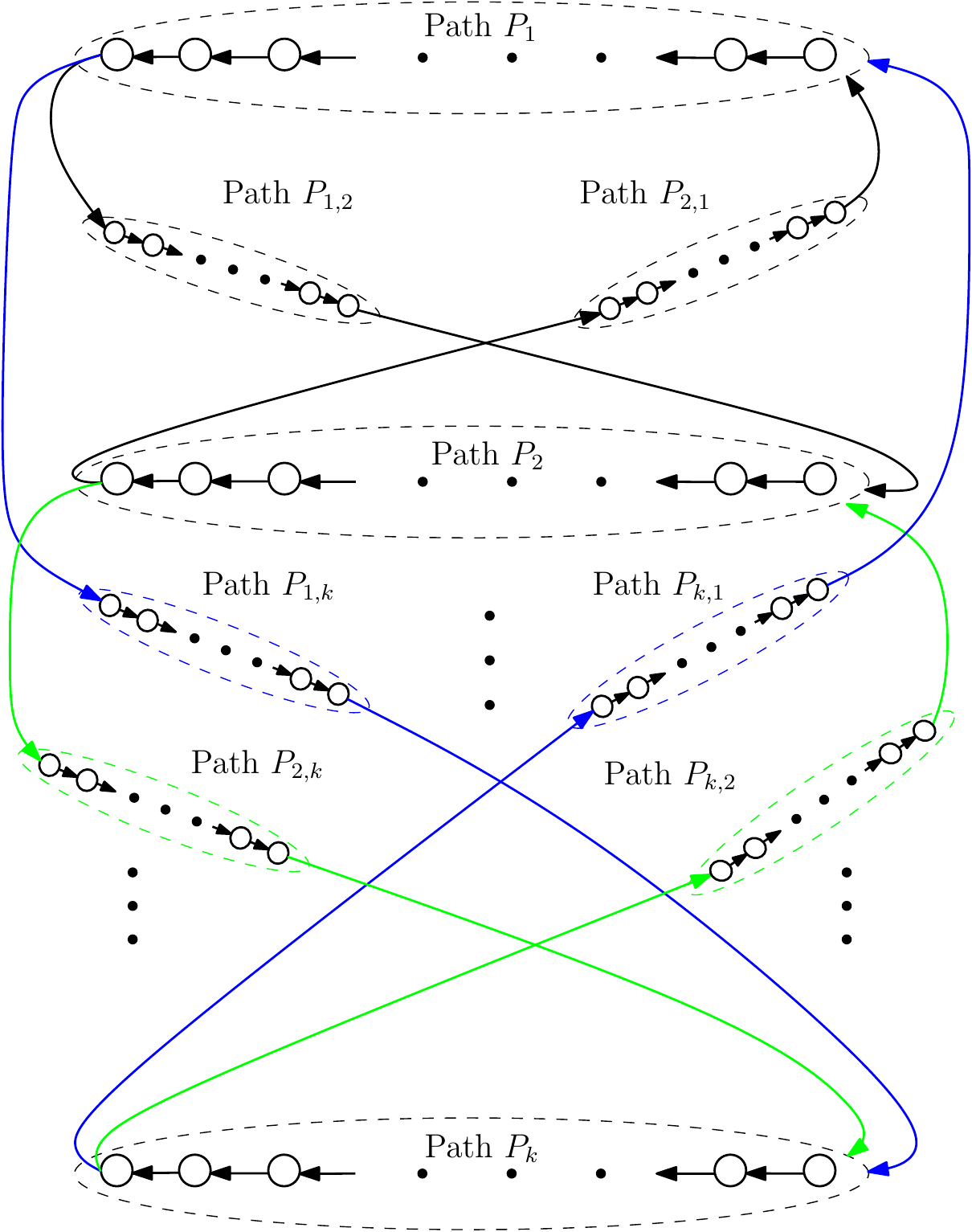}
 		\centering
 		\caption{An $ \mathsf{ICC} $ Digraph.}
 		\vskip-5pt
 		\label{figure1}
 		\end{figure}
\section{Construction of interlinked cycle cover ($ \mathsf{ICC} $)}

 	\subsection{Definition of $ \mathsf{ICC} $ digraphs}

We now construct a class of digraphs, which we call $ \mathsf{ICC} $ digraphs. In an $ \mathsf{ICC} $ digraph, there are two types of paths, Type-I and Type-II, where the terminal vertex of each Type-I path has an out-degree of $ k-1 $, and the terminal vertex of each Type-II path has an out-degree of $ 1 $, for some $k \geq 1$ (see Fig.~\ref{figure1}). More specifically, an $ \mathsf{ICC} $ digraph $D=(V,A)$ with $n$ vertices consists of
 	\begin{itemize}
 	\item $ k $ paths of Type-I, where $ k $ is a positive integer, 
 	\item $ k(k-1) $ paths of Type-II, and
 	\item interconnecting arcs in between paths of Type-I and Type-II, or between paths of Type-I.
 	\end{itemize}
 	
 	The $ k $ paths of Type-I are denoted by $ P_i $ for $ i=1,2,\dotsc,k $. Each $ P_i $ contains a sequence of $ n_i\geq 1 $ vertices $\{v_1^i,v_2^i,,\dotsc,v_{n_i}^i\}$, and arcs $ \{ (v_{a}^i\rightarrow v_{a+1}^i): \text{ for all } a=1,2,\dotsc,(n_i-1) \}$. 
 	 	
 	Similarly, the $ k(k-1) $ paths of Type-II are denoted by $ P_{i,j} $ for each ordered pair $ (i,j) $ from $ \{1,2,\dotsc,k\} $ where $ i\neq j $. Each $P_{i,j}$ contains a sequence of $ n_{ij} \geq 0 $ vertices $ \{ v_1^{ij},v_2^{ij},\dotsc,v_{n_{ij}}^{ij} \}$ and arcs $ \{(v_{a}^{ij}\rightarrow v_{a+1}^{ij}): \text{ for all } a=1,2,\dotsc,(n_{ij}-1)\} $. 
	  		
 	We now define interconnecting arcs between different paths: For each $(i,j)$, if $ n_{ij}\geq 1 $, one interconnecting arc connects the terminal vertex of $ P_i $ to the initial vertex of $ P_{i,j} $, i.e., $ (v_{n_i}^i\rightarrow v_1^{ij})\in A $, and another arc connects the terminal vertex of $ P_{i,j} $ to some vertex of $ P_j $, i.e., $ (v_{n_{ij}}^{ij}\rightarrow v_{q_i}^j)\in A $ for some $ v_{q_i}^j \in \{v_1^j,v_2^j,\dotsc,v_{n_j}^j\} $. Otherwise, ($ n_{ij}=0 $, i.e., $ P_{i,j}=\emptyset $), then one interconnecting arc connects the terminal vertex of $ P_i $ directly to some vertex of $ P_j $, i.e., $ (v_{n_i}^i\rightarrow v_{q_i}^j) \in A$. We require that the initial vertex $ v_1^j $ of each path $ P_j $ has at least one in-degree\footnote{We can show that our results also apply to the digraphs without this restriction.}. Fig. \ref{figure1} is a graphical representation of $ \mathsf{ICC} $ digraphs.
 	
The sets of vertices of all paths $ P_i $ and $ P_{i,j} $ are mutually disjoint. 
So, the total number of vertices in $D$ is
 	 \begin{align}
 	 	n=\sum\limits_{i} n_i+\sum\limits_{\substack{i,j\ \text{s.t.}\ i\neq j}} n_{ij}.
 	\end{align}
 		

Let  $x^i_a$ denote the message requested by receiver $v^i_a$, and $x^{ij}_a$ that requested by $v^{ij}_a$.

  	\subsection{Code construction for $ \mathsf{ICC} $ digraphs}
  	For any $ \mathsf{ICC} $ digraph $D$, we propose a valid index code that maps $n$ message packets (of $t$ bit each) to $\ell_\mathsf{ICC}(D)$ coded symbols (of $t$ bits each), consisting of
  	\begin{enumerate}
  	  	\item coded symbols obtained by the bitwise XOR ($ \oplus $) of each message pair requested by adjacent vertices of paths $ P_i $ for all $ i \in \{1,2,\dotsc,k\} $, and $n_i\geq 2$,
  	  	\begin{align}
  	  	\begin{split}
  	  	w_{a}^i=x_{a}^i \oplus x_{a+1}^i,\
  	  	\text{for}\ {a}=1,2,\dotsc,(n_i-1), \label{eqpi}
  	  	\end{split} 
  	  	\end{align}
  	  	(if $n_i=0$ or $1$, then no $w_{a}^i$  is constructed),
  	  	\item coded symbols obtained by the bitwise XOR of each message pair requested by adjacent vertices of paths $ P_{i,j}$ for all $ i\neq j,\ i,j \in \{1,2,\dotsc,k\} $, and $ n_{ij}\geq 2 $,
  	  	\begin{align}
  	  	w_{a}^{ij}=x_{a}^{ij} \oplus x_{a+1}^{ij},\ \text{for} \ a=1,2,\dotsc,(n_{ij}-1), \label{eqpij}
  	  	\end{align}
  	  	(if $ n_{ij}=0$ or $1$, then no $w_{a}^{ij}$ is constructed),
  	  	\item coded symbols obtained by the bitwise XOR of the message requested by the terminal vertex of $ P_{i,j}$ and that by $ v_{q_i}^j $ of $ P_j $ for all $ i\neq j$, 
  	  	\begin{align}
  	  	w_{n_{ij}}^{ij}=x_{n_{ij}}^{ij} \oplus x_{q_{i}}^{j}, \label{eqpnij}
  	  	\end{align}
  	  	(if $ n_{ij}=0$, then no $w_{n_{ij}}^{ij}$ is constructed), and
  	  	\item a coded symbol obtained by the bitwise XOR of messages requested by the terminal vertex of all paths $ P_i $,
  	  	\begin{align}
  	  	w'= \bigoplus \limits_{i=1}^k x_{n_i}^i. \label{leq}
  	  	\end{align}	
  	 \end{enumerate}

	\begin{remark}
	The encoding of the above code requires less than or equal to $t(n-1)$ bit-wise XOR operations.
	\end{remark}

  	Now, the index code constructed for the $ \mathsf{ICC} $ digraph is $ W=\{ (w_{a}^i)_{\forall i, \forall a},(w_{b}^{ij})_{\forall ij, \forall b},(w_{n_{ij}}^{ij})_{\forall ij},w'\} $. The total number of coded symbols, each of $ t $-bits, in $ W $ is,
  	\vskip-2pt	
  	\begin{align}
  	\ell_{\mathsf{ICC}}(D) &=1+\sum\limits_{i} (n_i-1)+ \sum\limits_{i,j\ \text{s.t.}\ i\neq j} (n_{ij}-1)+\sum\limits_{i,j\ \text{s.t.}\ i\neq j} 1 \nonumber \\
  	&=1+\sum\limits_{i} n_i+\sum\limits_{i,j\ \text{s.t.}\ i\neq j} n_{ij}-k=n-k+1.\label{ell}
  	\end{align}
   	\vskip-2pt	
  Let us show that all vertices in $ D$ can decode their respective requested messages from $ W $. From \eqref{eqpi}, in any path $ P_i $, all vertices $ v_{a}^i $, except the terminal vertex $ v_{n_i}^i $,  can decode their requested messages. This is because by construction, for each $ a=1,2,\dotsc,(n_i-1) $,  vertex $ v_{a}^i $ has message $ x_{a+1}^i $ as side information. 
  	
  From \eqref{eqpij}, in any path $ P_{i,j} $, all vertices $ v_{a}^{ij} $, except the terminal vertex $ v_{n_{ij}}^{ij} $, can decode their respective messages. This is because by construction, for all $ a=1,2,\dotsc,(n_{ij}-1) $, each vertex $ v_{a}^{ij} $ has message $ x_{a+1}^{ij} $ as side information. 

Similarly, $ v_{n_{ij}}^{ij} $ knows $ x_{q_{i}}^{j} $. Thus from \eqref{eqpnij} the terminal vertex of each $ P_{i,j} $ can decode its message. 
  	
  For $n_{ij}\geq 1$, and $i\neq j$, we evaluate the following:
  \begin{align}
    	& \bigoplus\limits_{a=q_i}^{n_{j}-1} w_{a}^j\oplus\bigoplus \limits_{b=1}^{n_{ij}-1} w_{b}^{ij} \oplus w_{n_{ij}}^{ij}  \nonumber\\
    	&=\bigoplus \limits_{a=q_i}^{n_{j}-1} (x_{a}^j \oplus x_{a+1}^j)\oplus\bigoplus \limits_{b=1}^{n_{ij}-1} (x_{b}^{ij} \oplus x_{b+1}^{ij})\oplus (x_{n_{ij}}^{ij} \oplus x_{q_{i}}^{j})  \nonumber\\
    	&=(x_{q_i}^j \oplus x_{n_j}^j)\oplus (x_{1}^{ij} \oplus x_{n_{ij}}^{ij})\oplus (x_{n_{ij}}^{ij} \oplus x_{q_{i}}^{j}) \nonumber\\ 
    	&=( x_{1}^{ij} \oplus x_{n_{j}}^{j})\triangleq w''_{ij}. \label{pathsummation}
  \end{align}  
  Similarly, we evaluate the following:
  \begin{align}
  &\bigoplus \limits_{\substack{h\in \{1,\dotsc,k\}\setminus\{i\}\\ \ \text{s.t.}\ n_{ih}=0}}\hskip-5pt \left(\bigoplus \limits_{a=q_i}^{n_h-1}w_{a}^h\right)
 =\bigoplus \limits_{\substack{h\in \{1,\dotsc,k\}\setminus\{i\}\\ \ \text{s.t.}\ n_{ih}=0}}\hskip-5pt \left(\bigoplus \limits_{a=q_i}^{n_h-1} (x_{a}^h\oplus x_{a+1}^h)\right) \nonumber \\
 & =\bigoplus \limits_{\substack{h\in \{1,\dotsc,k\}\setminus\{i\}\\ \ \text{s.t.}\ n_{ih}=0}} (x_{q_i}^h\oplus x_{n_h}^h)= Y_i \oplus Y'_i, \label{F1}
  \end{align}
 where,
 $Y_i\triangleq \bigoplus \limits_{\substack{h\in \{1,\dotsc,k\}\setminus\{i\}\\ \ \text{s.t.}\ n_{ih}=0}} x^h_{q_i} $, and $Y'_i \triangleq \bigoplus \limits_{\substack{h\in \{1,\dotsc,k\}\setminus\{i\}\\ \ \text{s.t.}\ n_{ih}=0}} x^h_{n_h} $.
 Again, we evaluate the following:
 \begin{align}
  \bigoplus \limits_{\substack{h\in \{1,\dotsc,k\}\setminus\{i\}\\ \ \text{s.t.}\ n_{ih}\geq 1}} \hskip-4pt w''_{ih}
  =\hskip-2pt\bigoplus \limits_{\substack{h\in \{1,\dotsc,k\}\setminus\{i\}\\ \ \text{s.t.}\ n_{ih}\geq 1}}\hskip-5pt (x_{1}^{ih}\oplus x_{n_h}^h)= Z_i \oplus Z'_i, \label{F2}
 \end{align} 
 where,
 $Z_i\triangleq\bigoplus \limits_{\substack{h\in \{1,\dotsc,k\}\setminus\{i\}\\ \ \text{s.t.}\ n_{ih}\geq 1}} x_{1}^{ih}$, and $Z'_i \triangleq \bigoplus \limits_{\substack{h\in \{1,\dotsc,k\}\setminus\{i\}\\ \ \text{s.t.}\ n_{ih}\geq 1}} x_{n_h}^{h}$. 
 On the other hand, we can expand $ w' $ as:
 \begin{align}
 w'=\bigoplus \limits_{i=1}^k x_{n_i}^i 
 &=x_{n_i}^i \oplus \bigoplus \limits_{\substack{h\in \{1,\dotsc,k\}\setminus\{i\}\\ \ \text{s.t.}\ n_{ih}=0}} x_{n_h}^h \oplus \bigoplus \limits_{\substack{h\in \{1,\dotsc,k\}\setminus\{i\}\\ \ \text{s.t.}\ n_{ih}\geq 1}} x_{n_h}^h \nonumber \\
 &=x_{n_i}^i \oplus Y'_i \oplus Z'_i. \label{F3}  
 \end{align}
 Now, using \eqref{F3}, \eqref{F1}, and \eqref{F2}, we evaluate the following: 
 \begin{align}
  (x_{n_i}^i\hskip-2pt \oplus\hskip-2pt Y'_i\hskip-2pt \oplus\hskip-2pt Z'_i)\oplus (Y_i\hskip-2pt \oplus\hskip-2pt Y'_i) \oplus (Z_i\hskip-2pt\oplus\hskip-2pt Z'_i)  
  = x_{n_i}^i\hskip-2pt \oplus\hskip-2pt Y_i\hskip-2pt \oplus\hskip-2pt Z_i. \label{terminalvertexdecode} 	 
  \end{align} 
  From \eqref{terminalvertexdecode}, the terminal vertex of each $P_i$, i.e., $v_{n_i}^i$, can decode its requested message $x_{n_i}^i$ because by construction, if $n_{ij}\geq 1$ (for the term $ Z_i $), then $v_{n_i}^i$ has $x_1^{ij}$ as side information, and if $n_{ij}=0$ (for the term $ Y_i$), then $v_{n_i}^i$ has $x_{q_i}^{j}$ as side information.  
  Therefore, from \eqref{eqpi}, \eqref{eqpij}, \eqref{eqpnij} and \eqref{leq} all the vertices in $D$ can decode their requested messages. Hence, the index code $ W $ is a \emph{valid index code}.
  \begin{definition}
  (\emph{Saved packets}) The term saved packets (or simply savings) is the number of packets saved (i.e., $ n-\ell_t(D) $) by sending coded packets (coded symbols) rather than sending uncoded message packets. 
  \end{definition}
  	
  	\begin{remark}
  	If $ k=1 $, then there exists only a single path $ P_1 $ in the $ \mathsf{ICC} $ digraph. Thus a valid index code in this case will be $ w_{a}^1=x_{a}^1 \oplus x_{a+1}^1$, for $a=1,2,\dotsc,(n_i-1) $, and $w'=x_{n_i}^1 $. 
        Here, the number of coded symbols equals the number of vertices, and so no saved packets is obtained.
       	\end{remark}
  	\section{Results}

\subsection{The $\mathsf{ICC}$ Scheme}

Now, we formally state our proposed $\mathsf{ICC}$ scheme:
\begin{definition}
 	   (\emph{Interlinked Cycle Cover} ($ \mathsf{ICC} $) \emph{scheme}) For any digraph, the $ \mathsf{ICC} $ scheme finds a set of disjoint $ \mathsf{ICC} $ subgraphs. It then (a) codes each of these $ \mathsf{ICC} $ subgraphs using the code construction described in Section III.B, and (b) sends uncoded messages requested by all remaining vertices (i.e., vertices which are not in any of these disjoint $ \mathsf{ICC} $ subgraphs). 
 	\end{definition}

  	     We denote an $\mathsf{ICC}$ digraph with $ k $ number of Type-I paths as a $k$-$\mathsf{ICC}$ digraph. Using the $\mathsf{ICC}$ scheme on an $\mathsf{ICC}$ digraph, we have the following:
 	\begin{Lemma}\label{theorem1}
 	For a $k$-$ \mathsf{ICC} $ digraph $D$ with $ t $-bit messages, for any $ k\geq 1 $ and any $ t\geq 1 $, the total number of saved packets using the $ \mathsf{ICC} $ scheme is $ k-1 $, i.e., $ n-\ell_{\mathsf{ICC}}(D)=k-1 $. 
 	\end{Lemma}
 	\hskip20pt\emph{Proof:}
 	Subtracting $ \ell_{\mathsf{ICC}}(D) $ of \eqref{ell} from $ n $ we get
 	\begin{equation}\label{nleq}
 	n-\ell_{\mathsf{ICC}}(D)=n-(n-k+1)=k-1. 	
 	\end{equation}

We can generalize this to an arbitrary digraph:
 	\begin{theorem} \label{corollary1}
 	For any digraph $ D $, a valid index code of length $ \ell_{\mathsf{ICC}}(D)=n-\sum_{i=1}^\psi (k_i-1) $ can be constructed using the $ \mathsf{ICC} $ scheme, where $ (k_i-1) $ is the saving in each disjoint $k_i$-$ \mathsf{ICC} $ subgraphs, and $ \psi $ is the number of disjoint $ \mathsf{ICC} $ subgraphs. 
 	\end{theorem}

 	\begin{IEEEproof}
        For any digraph $ D $ containing $ \psi $ number of disjoint $ \mathsf{ICC} $ subgraphs, each $ k_i$-$ \mathsf{ICC} $ subgraph gives a saving of $ k_i-1$ (Lemma~\ref{theorem1}), where $i \in \{1,\dotsc,\psi\}$. The total savings is the sum of savings in all disjoint $ \mathsf{ICC} $ subgraphs, i.e., $ \sum_{i=1}^{\psi}(k_i-1) $. Therefore, $ \ell_{\mathsf{ICC}}(D)=n-\sum_{i=1}^{\psi}(k_i-1) $. 
 	\end{IEEEproof}

\begin{remark}
The $\mathsf{ICC}$ subgraphs found by the $\mathsf{ICC}$ scheme are not unique. So, finding the best $\ell_\mathsf{ICC}(D)$ involves optimizing over all choices of disjoint $\mathsf{ICC}$ subgraphs in $D$.
\end{remark}

\subsection{ $ \mathsf{ICC} $ includes cycle cover and clique cover as special cases}

 	\begin{theorem}\label{lemma1}
 	The $ \mathsf{ICC} $ scheme includes the cycle cover scheme and the clique cover scheme as special cases. 
 	\end{theorem}	
 	\begin{IEEEproof}
 	Let us consider a cycle having $ L $ vertices and $(L-1) $ arcs for some integer $ L\geq 2 $, i.e., $ \{v_1,(v_1\rightarrow v_2), v_2, \dotsc,(v_{L_1-1}\rightarrow v_{L_1}),v_{L_1},(v_{L_1}\rightarrow v_{L_1+1}),v_{L_1+1},\dotsc,(v_{L-1}\rightarrow v_L),v_L,(v_{L}\rightarrow v_{1}),v_1\} $, where $ 1\leq L_1<L $. For this cycle, the cycle cover scheme provides a valid index code of length $ (L-1) $ \cite{neely,chaudhary}, i.e., 
 	\begin{align}
 	&(x_1\oplus x_2),(x_2\oplus x_3),\dotsc,(x_{L_1-1}\oplus x_{L_1}),{\color{red}(x_{L_1}\oplus x_{L_1+1})},\nonumber\\
 	&(x_{L_1+1}\oplus x_{L_1+2})\dotsc,(x_{L-2}\oplus x_{L-1}),(x_{L-1}\oplus x_L). \label{cycleeq}
 	\end{align}
 	Here the saving is always one packet. This cycle can be viewed as a $2$-$ \mathsf{ICC} $ digraph, which is shown in Fig. \ref{21}. Using the $ \mathsf{ICC} $ scheme we get a valid index code of length $ \ell_{\mathsf{ICC}}(D)=n-k+1=L-1 $, i.e., 
 	\begin{align}\label{icceq3}
 	&(x_1\oplus x_2),(x_2\oplus x_3),\dotsc,(x_{L_1-1}\oplus x_{L_1}),\nonumber \\ 
 	&(x_{L_1+1}\oplus x_{L_1+2}), (x_{L_1+2}\oplus x_{L_1+3}),\dotsc,(x_{L-1}\oplus x_L),\nonumber \\
        &{\color{red}(x_{L_1}\oplus x_L)}. 
 	\end{align}
 	Both the valid index codes from cycle cover and from $ \mathsf{ICC} $ are of the same length. The difference (indicated in red) is that $ (x_{L_1}\oplus x_{L_1+1}) $ does not appear in the $ \mathsf{ICC} $ code, and $ (x_{L_1}\oplus x_L) $ does not appear in the cycle-cover code. But one can generate $ (x_{L_1}\oplus x_{L_1+1}) $ from the existing code symbols of the $ \mathsf{ICC} $ codes and vice versa (the proof is straightforward). 
 	
        Furthermore, consider any digraph $ D $ with a total of $ n $ vertices and $ |C| $ disjoint cycles. The saving by cycle cover is one packet for each cycle. The messages of vertices not covered by these selected cycles are sent uncoded.  So, the total savings is the sum of savings for all disjoint cycles. The length of a valid index code from cycle cover is therefore
 	\begin{align}
 	\ell_{\mathsf{cyc}}(D)&=n-\sum_{r=1}^{|C|} 1 =n-|C|. \label{cycle}
 	\end{align}
 	Similarly, for the same digraph $ D $, considering each cycle as a $2$-$ \mathsf{ICC} $ subgraph, Theorem~\ref{corollary1} gives
 	\begin{align}
	\ell_{\mathsf{ICC}}(D)&=n-\hskip-2pt\sum_{i=1}^{|C|}\hskip-2pt(k_i-1) = n-\sum_{i=1}^{|C|}(1)=n-|C|. \label{icc}
	\end{align}
 	Hence, both schemes return the same index code length for any digraph $ D $, if the $ \mathsf{ICC} $ scheme assigns one $ \mathsf{ICC} $ subgraph to each disjoint cycle. Moreover, the index codes from both schemes are equivalent (using the above argument). This proves that cycle cover is also a special case of $ \mathsf{ICC} $.
 	\begin{figure}[t]
 	        \centering
 	        \subfigure[]{
 	                \includegraphics[height=3.2cm,keepaspectratio]{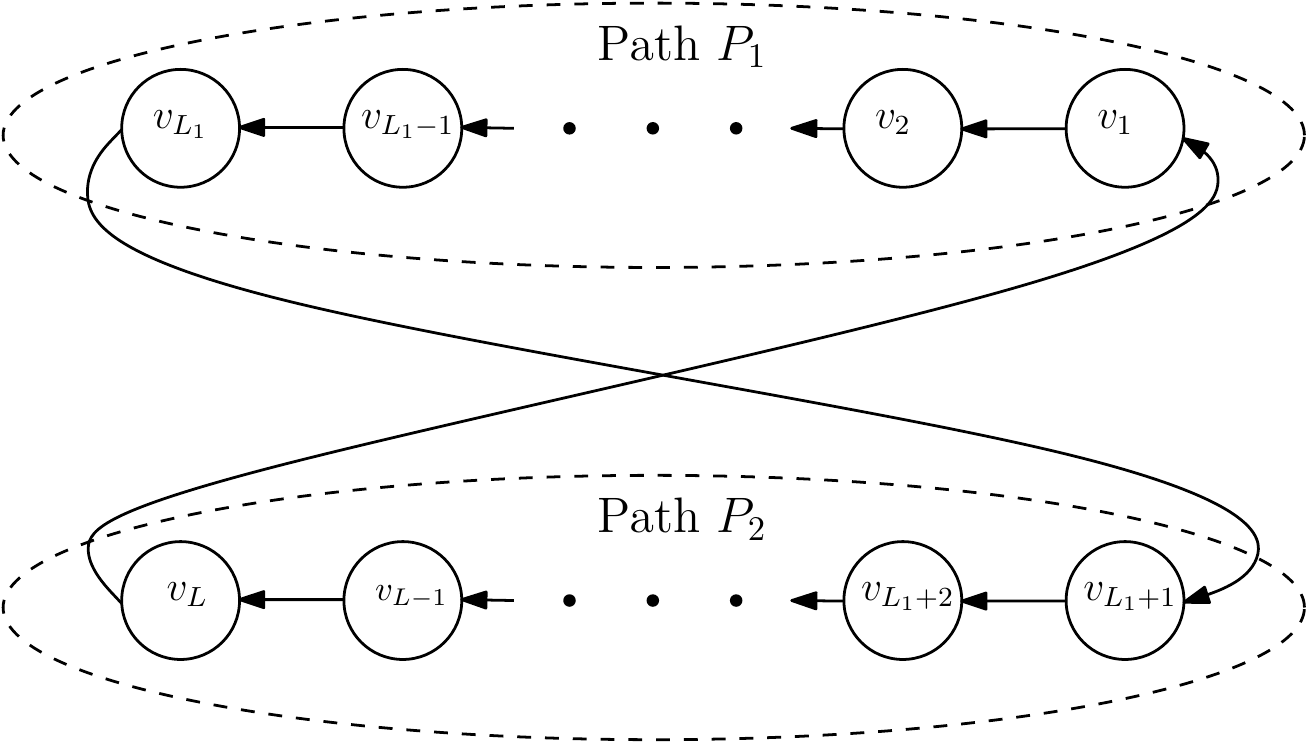}
 	               \label{21}                 
 	        }
 	        \hskip-6pt
 	        ~ 
 	        \subfigure[]{
 	                \includegraphics[height=3.3cm,keepaspectratio]{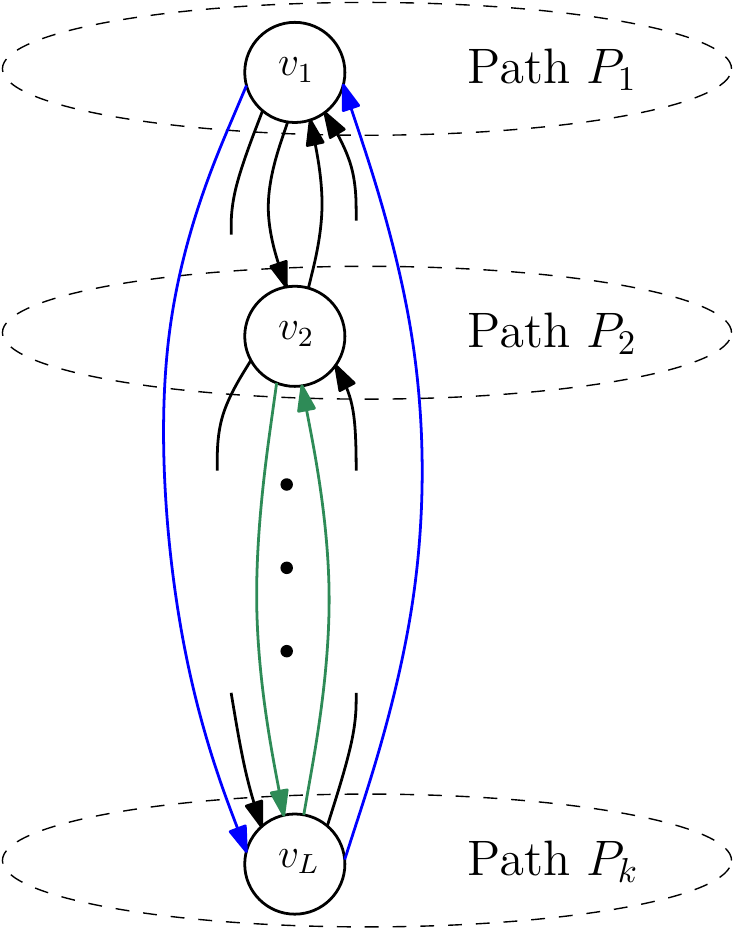}
 	                \label{22}	                
 	        }
 	       \vskip-2pt
 	        \caption{Special cases of $ \mathsf{ICC} $ digraphs (a) with $ k=2,\ n_1=L_1,\ n_2=L-L_1,\ n_{12}=n_{21}=0,\ v_{q_2}^1=v_1,\ \text{and}\ v_{q_1}^2=v_{L_1+1}$ forming a cycle , and (b) with any $ k=L\geq 1,\ n_i=1\ \forall i,\ \text{and} \ n_{ij}=0\ \forall i\neq j $ forming a clique.}
 	       \vskip-6pt
 	\end{figure}
 	
 	To prove clique cover as a special case of $ \mathsf{ICC} $, let us consider a clique of $ L $ vertices $\{v_1,\dotsc, v_L\}$ where, $ L\geq 1 $. The valid index code for this clique using the clique cover scheme is of length one, i.e., $ (x_1\oplus x_2\oplus \dotsc \oplus x_L) $. The clique can be viewed as a $L$-$ \mathsf{ICC} $ digraph, which is shown in Fig. \ref{22}. The $ \mathsf{ICC} $ scheme gives the same valid index code as that given by clique cover. 

        Furthermore, consider any digraph $ D $ with $ n $ vertices and $ |\chi| $ disjoint cliques, where each clique $r \in \{1,2,\dotsc,|\chi|\}$ consists of $ n_{r} $ vertices. The saving by clique cover is $ n_{r}-1 $ packets for each clique $r$. The messages corresponding to vertices not covered by these disjoint cliques are sent uncoded. So, the total savings is the sum of savings for each disjoint clique. The length of a valid index code from clique cover is
 	 	\begin{align}
 	 	\ell_{\mathsf{cc}}(D)&=n - \sum_{r=1}^{|\chi|}( n_{r} - 1) =n-\sum_{r=1}^{|\chi|}n_{r}+|\chi|. \label{clique}
 	 	\end{align}
 	 	Similarly, for the same digraph $ D $, considering each clique as a $n_i$-$\mathsf{ICC} $ subgraph, the length of a valid index code by the $ \mathsf{ICC}$ scheme using Theorem~\ref{corollary1} is
 	 	\begin{align}
 		 \ell_{\mathsf{ICC}}(D)&=n-\sum_{i=1}^{|\chi|}(n_i-1) 
 		 =n-\sum_{i=1}^{|\chi|}n_i+|\chi|. \label{cc}
 		\end{align}
 	 	Hence, both schemes return the same index code length for any digraph $ D $, if the $ \mathsf{ICC} $ scheme assigns one $ \mathsf{ICC} $ digraph to each disjoint clique. Moreover, the index codes from both schemes are equivalent. This proves that clique cover is a special case of $ \mathsf{ICC} $.  	 	 	 
 	\end{IEEEproof}	

\subsection{ $ \mathsf{ICC} $ is optimal for any $ \mathsf{ICC} $ digraph}

We first prove the following lemma:

 	\begin{Lemma}\label{lemma2}
	 In an $ \mathsf{ICC} $ digraph, any cycle that contains a vertex $ v\in P_i $ must also contain the terminal vertex $v_{n_i}^i$, and any cycle that contains a vertex $ v\in P_{i,j} $ must also contain the terminal vertex $ v_{n_j}^j $.  
 	\end{Lemma}
 	\begin{IEEEproof}
         For any cycle containing $v$, there must be a path, say $P$, from $v$ back to itself.

         (Case 1) If $v \in P_i$ (where $P_i$ is not a cycle), then the path $P$ must leave $P_i$. By construction, any arc that leaves $P_i$ originates from $v_{n_i}^i$. Hence, $P$ must contain $v_{n_i}^i$. So, any cycle that contains $v \in P_i$ must also contain $v_{n_i}^i$.

         (Case 2)  If $v \in P_{i,j}$ (where $P_{i,j}$ is again not a cycle), then the path $P$ must leave $P_{i,j}$. There is only one arc leaving $P_{i,j}$, which is from $v_{n_{ij}}^{ij} \in P_{i,j}$ to $v_{q_i}^j \in P_j$. Note that $v \notin~P_j$. Repeating the argument for Case~1, the path $P$ must go through $v_{n_j}^j$ before going back to $v$ (to form a cycle). So any cycle that contains $v \in P_{i,j}$ must also contain $v_{n_j}^j$.
 	\end{IEEEproof}

        With the above lemma, we now show the following:

 	\begin{theorem}\label{theorem2}
 	For any $ t \geq 1$, the linear index code given by the $ \mathsf{ICC} $ scheme is optimal for any $ \mathsf{ICC} $ digraph, i.e., $\ell^*_t(D)=\ell_{\mathsf{ICC}}(D)$. 
 	\end{theorem}
 	\begin{IEEEproof}
         It has been shown~\cite{maisbound} that for any digraph $D$ and any message length $t$, $\ell_t^*(D) \geq \mathsf{MAIS}(D)$, where $\mathsf{MAIS}(D)$ is the order of a maximum acyclic induced subgraph of $D$. To obtain $ \mathsf{MAIS}(D) $, one has to remove the minimum number of vertices from $D$ to make it acyclic.

        Consider a $k$-$\mathsf{ICC}$ digraph $D$. From Lemma~\ref{lemma2}, we know that any cycle must contain the terminal vertex of a Type-I path, say $v_{n_i}^i$.  Note that any outgoing arc from $v_{n_i}^i$ terminates at a vertex in either (a) $P_{i,j}$ for some $j$, or (b) $P_j$ for some $j \neq i$. Using the same argument in the proof of Lemma~\ref{lemma2}, any cycle that contains $v_{n_i}^i$ must also contain $v_{n_j}^j$ for some $j \neq i$. So, every cycle must contains at least two terminal vertices of Type-I. Therefore, removing $(k-1)$ terminal vertices of Type-I paths makes $D$ acyclic. 
This gives, $\mathsf{MAIS}(D) \geq n- (k-1)$.
		 	
	The removal of any $ k-2 $ or fewer vertices from an $ \mathsf{ICC} $ digraph cannot make the digraph acyclic. This can be proved by the following reasoning. The removal of any vertex, say $v$ (which must belong to some path $ P_i $ or path $ P_{j,i} $), to break cycles containing $v$ is no better than the removal of $v_{n_i}^i$ (which also breaks those cycles). This is due to Lemma~\ref{lemma2}. It follows that the removal of any $ k-2 $ or fewer vertices cannot be better than the removal of $ k-2 $ or fewer terminal vertices.  Even if $ k-2  $ terminal vertices are removed, say $\{v_{n_i}^i: i = 3, 4, \dotsc, k\}$ without loss of generality, $P_1$, $P_{1,2}$, $P_2$, and $P_{2,1}$ form a cycle, which is not removed.  Thus, $ k-1 $ is the least possible removal to make an $ \mathsf{ICC} $ digraph acyclic, i.e., $\mathsf{MAIS}(D) \leq n - (k-1)$.

Combining the upper and lower bounds, we have
 	\begin{equation}
 	\mathsf{\mathsf{MAIS}}(D) =n-k+1 \leq \ell_t^*(D). \label{ell8}
 	\end{equation}   
 	From Lemma~\ref{theorem1} we get,
 	\begin{equation}
 	\ell_{\mathsf{ICC}}(D)=n-k+1 \geq \ell_t^*(D). \label{ellICC} 
 	\end{equation}
 	From \eqref{ell8} and \eqref{ellICC}, we get $\ell^*_t(D) = \ell_\mathsf{ICC}(D) $. 
 	\end{IEEEproof}

    For any $\mathsf{ICC}$ digraph $D$, $\beta_t(D) = \ell_\mathsf{ICC}(D) = n-k+1$, which is independent of $t$. This means $\beta(D) = \inf_t \beta_t(D) = n-k+1 = \ell_\mathsf{ICC}(D)$, and we have the following:
     \begin{corollary}
      For any $\mathsf{ICC}$ digraph, the $\mathsf{ICC}$ scheme achieves $\beta(D)$.
    \end{corollary}
        
\subsection{ $ \mathsf{ICC} $ can outperform existing techniques }
 	For some digraphs, $ \mathsf{ICC} $ can outperform existing techniques such as clique cover ($ \mathsf{cc} $) \cite{ISCOD}, fractional clique cover ($ \mathsf{fcc} $) \cite{linearprogramming}, partial clique cover ($ \mathsf{pcc} $) \cite{ISCOD}, fractional partial clique cover ($ \mathsf{fpcc} $) \cite{localtimesharing}, cycle cover ($ \mathsf{cyc} $) \cite{maisbound,neely,chaudhary}, local chromatic number ($ \mathsf{lc} $) \cite{localgarphcoloring}, and local time sharing bounds $ (b(\mathscr{R}_{\mathsf{LTS}}(D)) \ \text {and} \ b_{\mathsf{LTS}}(D))$ \cite{localtimesharing}. Here are two examples: 
  \begin{figure}[t]
   	        \centering
   	        \subfigure[]{
   	                \includegraphics[height=3cm,keepaspectratio]{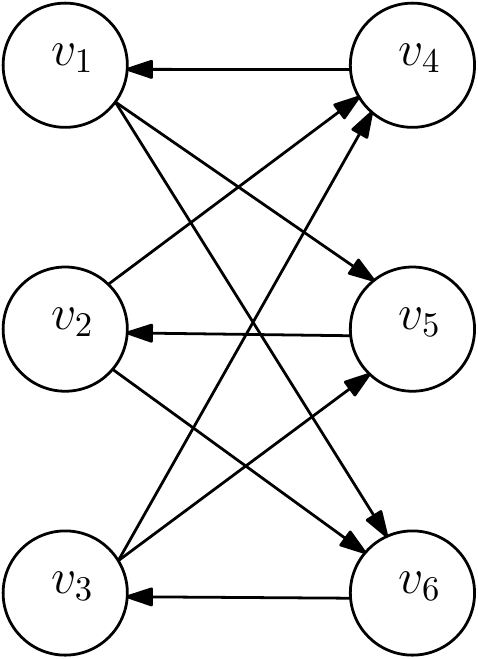}
   	               \label{exp2}                
   	        }
   	        \qquad
   	        ~ 
   	        \subfigure[]{
   	                \includegraphics[height=3cm,keepaspectratio]{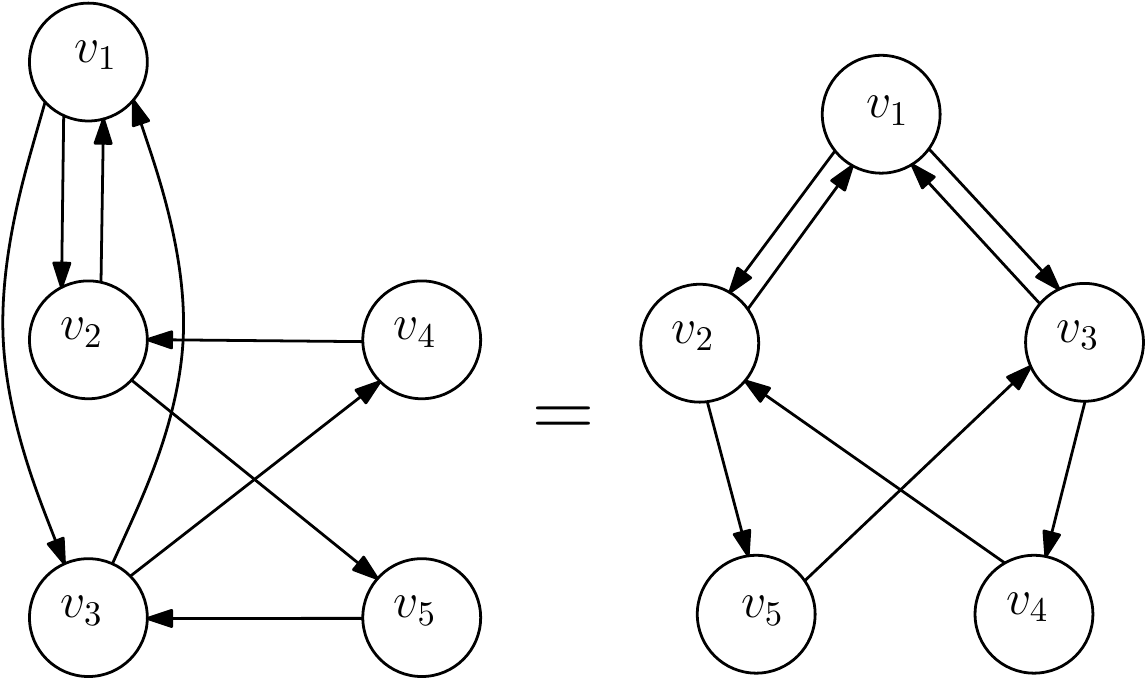}
   	                \label{exp1}                
   	        }
   	       \vskip-2pt 
   	       \caption{$ \mathsf{ICC} $ digraphs: (a) $ D_1 $ with $ n=6 $, $  k=3 $, $ n_1=n_2=n_3=2 $ and all $ n_{ij}=0$, and (b) $D_2 $ with $ n=5 $, $  k=3 $, $ n_1=1,\ n_2=n_3=2 $, and all $ n_{ij}=0$.}
   	       \vskip-6pt
   \end{figure}
  	
 	For the $\mathsf{ICC}$ digraph $ D_1 $ in Fig. \ref{exp2}, a valid index code from the $ \mathsf{ICC} $ scheme is $ \{ x_4\oplus x_1, x_5\oplus x_2, x_6\oplus x_3, x_1\oplus x_2\oplus x_3 \} $, which is of length four i.e. $ \ell_{\mathsf{ICC}}(D_1)=4 $. For this digraph, $ \beta(D_1)=\ell_{\mathsf{ICC}}(D_1)=4 < \ell_\mathsf{fpcc}(D_1)=4.5  < \ell_\mathsf{lc}(D_1)= \ell_\mathsf{pcc}(D_1)= \ell_\mathsf{cyc}(D_1)=5 < \ell_\mathsf{cc}(D_1)=\ell_\mathsf{fcc}(D_1)=6 $. 
 	
 	Similarly, for the $\mathsf{ICC}$ digraph $ D_2 $ in Fig. $ \ref{exp1} $, a valid index code from the $ \mathsf{ICC} $ scheme is $ \{ x_4\oplus x_2, x_5\oplus x_3, x_1\oplus x_2\oplus x_3 \} $, which is of length three i.e. $ \ell_{\mathsf{ICC}}(D_2)=3 $. For this digraph, $ \beta(D_2) =\ell_{\mathsf{ICC}}(D_2)=3 < b_\mathsf{LTS}(D_2)=b(\mathscr{R}_{\mathsf{LTS}}(D_2))=7/2$  $<\ell_\mathsf{lc}({D_2})=4 $. 

        Furthermore, some of the existing techniques (e.g., $\mathsf{pcc}$, $\mathsf{lc}$) use maximum distance separable (MDS) codes, which requires $t$ to be sufficiently large.

We now describe a class of digraphs where the $\mathsf{ICC}$ scheme outperforms the local chromatic number in the order of the order of the digraph (i.e., the number of vertices).
Consider a digraph $ D $ with even number of vertices,  $n=2k$, where $ k $ is any positive integer. Furthermore, the vertices can be grouped into two sets, without loss of generality, say $ V_1=\{v_1,\dotsc,v_k\} $ and $V_2=\{v_{k+1},\dotsc,v_n\}  $, such that for each $ i \in \{1,\dotsc,k\} $, $ v_{k+i} $ knows a message requested by $ v_i $, and $ v_i $ knows messages requested by all $ V_2 \setminus \{v_{k+i}\} $. We can show that the gap $ \ell_{\mathsf{lc}}(D)~- \ell_{\mathsf{ICC}}(D) $ for this type of digraphs grows linear with $ n $. Note that $ D_1 $ in Fig.~\ref{exp2} belongs to this class of digraphs with $k=3$.

\section{Conclusion}
For unicast index coding problems, we designed a new coding scheme, called interlinked cycle cover ($ \mathsf{ICC} $), which exploits  interlinked cycles in the digraph. Our proposed $\mathsf{ICC}$ scheme includes clique cover and cycle cover as special cases. We proved that this scheme gives an optimal index code for a class of digraphs, namely, $ \mathsf{ICC}  $ digraphs, and it can outperform existing schemes.  


\bibliographystyle{IEEEtran}
\bibliography{myref}

\end{document}